%% file: _main.tex
\documentclass[conference]{IEEEtran}

\usepackage{subcaption}

\usepackage{adjustbox}
\usepackage{amsmath,amssymb,amsfonts}
\usepackage{color,soul}
\usepackage{enumitem}
\usepackage{listings}
\usepackage{makecell}
\usepackage{mathtools}
\usepackage{multirow}
\usepackage{pifont}
\usepackage{textcomp}
\usepackage{url}
\setlist[itemize]{leftmargin=*} \usepackage{algorithm,algorithmic}
\def\BibTeX{{\rm B\kern-.05em{\sc i\kern-.025em b}\kern-.08em
    T\kern-.1667em\lower.7ex\hbox{E}\kern-.125emX}}
\usepackage{graphicx}
\usepackage{textcomp}
\usepackage{xcolor}
\usepackage{booktabs}
\usepackage{tabularx,hhline}
\usepackage{tikz}

\usepackage[strings]{underscore}
\usepackage[left=1.62cm,right=1.62cm,top=0.7in]{geometry}
\usepackage{float}
\usepackage{cite}
\usepackage[hidelinks]{hyperref}
\begin{document}

\title{Cyber-Physical Co-Simulation of Load Frequency Control under Load-Altering Attacks}
\IEEEaftertitletext{\vspace{-12\baselineskip}}

\author{
    \IEEEauthorblockN{
        \textbf{Michał Forystek}\IEEEauthorrefmark{1},
        \textbf{Andrew D. Syrmakesis}\IEEEauthorrefmark{2}, \textbf{Alkistis Kontou}\IEEEauthorrefmark{2}, \textbf{Panos Kotsampopoulos}\IEEEauthorrefmark{2},\\ \textbf{Nikos D. Hatziargyriou}\IEEEauthorrefmark{2}, 
        \textbf{Charalambos Konstantinou}\IEEEauthorrefmark{1}
    }
    
    \IEEEauthorblockA{\IEEEauthorrefmark{1}CEMSE Division, King Abdullah University of Science and Technology (KAUST)}
    \IEEEauthorblockA{\IEEEauthorrefmark{2}School of Electrical and Computer Engineering, National Technical University of Athens}
    \vspace{-10mm}
}

\maketitle

\input{0_abstract}
\input{1_introduction}
\input{2_model}
\input{3_implementation}
\input{4_results}
\input{5_conclusion}
\vspace{-2mm}
\input{acknowledgments}
\vspace{-2mm}
\bibliographystyle{ieeetr}
\bibliography{references}

\vfill

\end{document}

%% file: 0_abstract.tex
\begin{abstract}

Integrating Information and Communications Technology (ICT) devices into the power grid brings many benefits. However, it also exposes the grid to new potential cyber threats. Many control and protection mechanisms, such as Load Frequency Control (LFC), responsible for maintaining nominal frequency during load fluctuations and Under Frequency Load Shedding (UFLS) disconnecting portion of the load during an emergency, are dependent on information exchange through the communication network. The recently emerging Load Altering Attacks (LAAs) utilize a botnet of high-wattage devices to introduce load fluctuation. In their dynamic form (DLAAs), they manipulate the load in response to live grid frequency measurements for increased efficiency, posing a notable threat to grid stability. Recognizing the importance of communication networks in power grid cyber security research, this paper presents an open-source co-simulation environment that models the power grid with the corresponding communication network, implementing grid protective mechanisms. This setup allows the comprehensive analysis of the attacks in concrete LFC and UFLS scenarios.

\end{abstract}
\begin{IEEEkeywords}
Load altering attacks, load frequency control, digital twin, real-time digital simulation.
\end{IEEEkeywords}
 

%% file: 1_introduction.tex
\vspace{-3mm}
\section{Introduction}
\label{sec:Introduction}
\vspace{-1mm} 

The digitization of power systems has accelerated the integration of Information and Communication Technology (ICT) devices into the grid infrastructure. While these advancements enhance monitoring, automation, and control capabilities, they simultaneously expand the grid’s cyber-physical attack surface \cite{CyberPhysicalInterdependence, CybersecurityInPowerGrids, CyberPhysicalSecurity}. Many critical control and protection functions, such as Load Frequency Control (LFC), which adjusts the generators' load setpoint to restore nominal frequency and Under Frequency Load Shedding (UFLS), which disconnects portion of the load during emergency conditions, depend heavily on timely and reliable information exchange through communication network \cite{RobustPowerSystemFrequencyControl, AComprehensiveReviewOfLFC}. This tight coupling between the cyber and physical layers necessitates new methodologies for assessing the grid’s resilience under cyber threats.

Among the emerging classes of cyber-physical threats, Load Altering Attacks (LAAs) have drawn significant attention due to their ability to cause large-scale disturbances without targeting traditional control centers or substations \cite{BlackIoT, surveyLAA, GridShock}. These attacks exploit the abundance of high-wattage, IoT-controlled devices, such as electric vehicle chargers, heat pumps, and HVAC systems, by forming botnets capable of manipulating aggregated load in a coordinated fashion. While static LAAs (SLAAs) are characterized by prescheduled disruptions, dynamic LAAs (DLAAs) adjust their behavior in real-time based on grid frequency measurements, effectively embedding a control loop into the adversary’s strategy \cite{DynamicLAA}. Finally, the Measurement-based DLAA (MDLAA) predicts the attack vector based solely on the frequency measurements contrary to DLAA, which requires the grid topology knowledge \cite{MDLAA}. These attacks challenge the conventional assumptions about the location and detectability of malicious behavior within the grid.

The interaction between malicious load changes and the grid’s frequency control mechanisms introduces complex dynamics. LFC, in particular, continuously adjusts generator output to maintain nominal frequency in the presence of load changes. When an adversary manipulates load based on live frequency measurements, the attack can act as an anti-control system, deliberately opposing the stabilizing actions of LFC \cite{DynamicLAA, MDLAA}. Additionally, the communication network disruptions, such as latency or packet loss, can influence the performance of control mechanisms \cite{LFCDelays1, LFCDelays2, 10518173}.

Co-simulation offers a solution to capture the intertwined cyber and physical effects by coupling real-time digital grid simulators with communication network emulators. Such environments allow researchers to evaluate realistic cyber-physical scenarios, including attacks that operate across both domains. Prior works have explored co-simulation frameworks combining tools such as RTDS, OPAL-RT, NS-3, EXata CPS, Python, and Mininet \cite{PolitecnicoDiTorino, RT-METER, RealTimeFederated, forystek2025exploring}. However, most efforts either focus on specific protection schemes or lack an open-source implementation tailored to cyber-physical threat evaluation.

This work presents a practical, open-source co-simulation environment that integrates an industry-grade real-time digital simulator, RTDS, with Containernet emulator, the fork of Mininet with native Docker support, to model cyber-physical interactions under LAA scenarios. In this context, LFC and UFLS mechanisms are implemented within the power model, while the attack logic and communication protocols are executed within the network emulation part. The environment is designed to support experimentation under varying network conditions and adversary knowledge levels, enabling robust cybersecurity studies of grid operations.

While previous work has studied LFC and LAAs independently, few have investigated their interaction in a real-time co-simulation context. Moreover, existing DLAA and MDLAA studies \cite{DynamicLAA, MDLAA} do not account for communication network presence or use real-time simulation platforms. Similarly, co-simulation works typically do not offer publicly available frameworks. To address these gaps, this paper contributes the following:
\begin{itemize}
    \item An open-source communication network emulation designed to integrate with the RTDS simulator, supporting multiple types of LAA scenarios\footnote{Co-simulation network emulation and RTDS models available on GitHub at \cite{RTDSModelsGithub, CoSimulationGithub}.}.
    \item A comparative study of LFC and UFLS performance under SLAA and DLAA attacks.
    \item An evaluation of effects of altered networking conditions on the effectiveness of LFC and UFLS under different variants of LAAs.
\end{itemize}

%% file: 2_model.tex
\vspace{-1mm}
\section{Models of LFC and LAAs}
\label{sec:ModelMethodology}
\vspace{-1mm}

This section presents analytical models of LFC and LAA variants, SLAA, DLAA, and MDLAA. The state-space representation of LAAs enables efficient software implementation. The LFC model visualization, through a block diagram of transfer functions, provides a  presentation of control dynamics.

\vspace{-1mm}
\subsection{LFC Modeling}
\label{subsec:ModelOfLoadFrequencyControl}
\vspace{-1mm}

\begin{figure}[t]
    \centering
    \includegraphics[width=\linewidth]{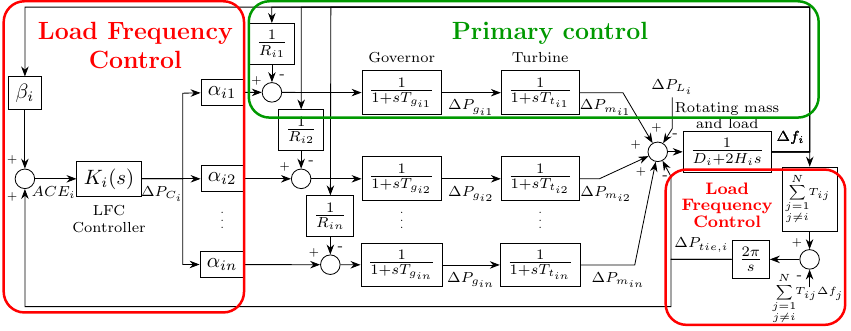}
    \vspace{-2mm}
    \caption{Block diagram of the power system area implementing LFC.}
    \vspace{-5mm}
    \label{fig:area_schema_full}
\end{figure}

To visualize the system implementing the primary and secondary control (LFC) in transmission networks, we use the area-level block diagram of frequency control \cite{RobustPowerSystemFrequencyControl}, as illustrated in Fig. \ref*{fig:area_schema_full}. The transfer functions are used to represent different components of the system in each area $i \in \{1,2,\ldots,N\}$. The upper-right part of the diagram depicts the primary control for each generator, achieved by multiplying $\Delta f_i$ by the droop gain $R_i$. This control type responds to local frequency changes, adjusting the generator's mechanical output within seconds. On the other hand, the LFC loop, shown as the leftward input to the governor, adjusts the load reference setpoint of participating generators to restore nominal frequency. Primary control parameters are specific to each generator, while the secondary control operates at area-level, distributing corrective actions among the $n$ LFC-driven generators according to their respective weighted factor $\alpha_{ij}$, where $j\! \in\! \{1,2,\ldots,n\}$. Detailed state-space representation of LFC is available in \cite{forystek2025exploring, 10349957}.

\vspace{-1mm}
\subsection{LAA Modeling}
\label{subsec:ModelOfStaticAndDynamicLAA}
\vspace{-1mm}

SLAAs can be modeled as a set of differential equations \cite{lakshminarayana2022load}. In this case, as shown in \eqref{eq:staticLAA}, the system's internal state is modeled as a concatenation of voltage phase angles of generator buses, voltage phase angles of load buses, and rotor frequency deviation of generator buses, vectors $\delta$, $\theta$, and $\omega$ respectively. System input is defined by vector $P^{LS}$ as a secure portion of the load at each bus. In contrast, vector $\epsilon^L$ represents the vulnerable portion of the load used to perform SLAA.

\small
\vspace{-1mm}
\begin{equation}
\label{eq:staticLAA}
    E \begin{bmatrix}
        \dot{\delta} \\ 
        \dot{\theta} \\
        \dot{\omega}
    \end{bmatrix} = A \begin{bmatrix}
        \delta \\
        \theta \\
        \omega
    \end{bmatrix} + B(P^{LS} + \epsilon^L)
\end{equation}
\vspace{-2.5mm}
\normalsize

The \eqref{eq:staticLAAMatrixAAndMatrixB}, and \eqref{eq:imagAdmittanceMatrix} show the system, input, and mass matrices $A$, $B$, and $E$. We define $A^{\kern-0.1em 1} \!\!:=\! diag(\!A\!*\!\mathbf{1}\!)$ where $\mathbf{1}$ is the column vector of ones. The generator inertias, damping coefficients, and proportional and integral coefficients for primary and secondary control are shown by the diagonal matrices $M$, $D^G$, $K^P$, and $K^I$. The admittance matrix $H_{bus}$ \eqref{eq:imagAdmittanceMatrix} depicts connections between generator-to-generator ($H^{GG}$), generator-to-load bus ($H^{GL}$), load bus-to-generator ($H^{LG}$), and load bus-to-load bus ($H^{LL}$). If two buses are unconnected, the respective matrix element equals zero. Finally, $I$ is the identity matrix of the appropriate dimensions.

\small
\vspace{-3mm}
\begin{equation}
\label{eq:staticLAAMatrixAAndMatrixB}
    \setlength{\arraycolsep}{2pt}
    A = \begin{bmatrix}
        0 & 0 & I \\
        -\!H^{\!LG} & H^{\!LG^1}\kern-0.6em+\!\!H^{\!LL^1}\kern-0.7em-\!\!H^{\!LL} & 0 \\
        K^{\!I}\kern-0.4em+\!\!H^{\!GG^1}\kern-0.7em-\!\!H^{\!GG}\kern-0.4em+\!\!H^{\!GL^1} & -\!H^{\!GL} & K^{\!P}\kern-0.5em+\!\!D^{\kern-0.07emG} 
    \end{bmatrix}
\end{equation}
\vspace{-2mm}
\begin{equation}
    \label{eq:imagAdmittanceMatrix}
    E = \begin{bmatrix}
        I & 0 & \kern-0.5em 0 \\
        0 & 0 & \kern-0.5em 0 \\
        0 & 0 & \kern-0.5em -M
    \end{bmatrix};\quad
     B = \begin{bmatrix}
        0 \\
        I \\
        0
    \end{bmatrix}; \quad
    H_{bus} = \begin{bmatrix}
        H^{GG} H^{GL} \\
        H^{LG} H^{LL}
    \end{bmatrix}
\end{equation}
\vspace{-2mm}
\normalsize

The previous model effectively represents SLAA, however, this attack does not alter the system stability, which is a key distinction of DLAA \cite{DynamicLAA}. To model DLAA, first we modify \eqref{eq:staticLAA} by substituting $\theta$ \eqref{eq:theta} to then obtain the non-descriptor form. Next, we incorporate the attack into the system matrix to affect stability. As shown in \eqref{eq:stateSpaceDynamic}, the DLAA proportional coefficient vector $K^{LG}>\nolinebreak0$ indicates manipulation of the vulnerable load portion that can modify the system state during an attack.

\small
\vspace{-3mm}
\begin{equation}
    \label{eq:theta}
    \theta = H^{inv}(H^{LG}\delta\! - \!P^L); \quad H^{inv}\kern-0.2em = (H^{LG^1}\kern-0.4em+\!H^{LL^1}\kern-0.5em-\!H^{LL})^{-1}
\end{equation}
\begin{equation}
    \label{eq:stateSpaceDynamic}
   \begin{bmatrix}
        \dot{\delta} \\ \dot{\omega}
    \end{bmatrix} = A'\begin{bmatrix}
        \delta \\ \omega
    \end{bmatrix} + B'\left(\begin{bmatrix}
        0 & \kern-0.4em -K^{LG}
    \end{bmatrix}\begin{bmatrix}
        \delta \\ \omega
    \end{bmatrix}+P^{LS}\right)
\end{equation}
\vspace{-2mm}
\normalsize

These changes influence the system matrix as shown in \eqref{eq:matrixADynamicNew}. We obtain the new system matrix $A^*$ \eqref{eq:matrixA*Dynamic} of the system under attack. The new form of the system state-space equations is shown in \eqref{eq:stateSpaceDLAAFinal}. This model integrates the DLAA into the system matrix, meaning that changes to $K^{LG}$ can affect system stability by shifting the eigenvalues of $A^*$.

\small
\vspace{-1.5mm}
\begin{equation}
    \label{eq:matrixADynamicNew}
    A^* = A' + B'\begin{bmatrix}
        0 & \kern-0.4em -K^{LG}
    \end{bmatrix}
\end{equation}
\vspace{-3mm}
\begin{align}
    \label{eq:matrixA*Dynamic}
    &\begin{aligned}
        A^*\! &=\! \left[\begin{matrix}
        0 \\
        M^{-1}(H^{GG}\kern-0.4em-\!\!H^{GG^1}\kern-0.6em-\!\!H^{GL^1}\kern-0.5em+\!\!H^{GL}H^{inv}H^{\!LG}\kern-0.3em-\!\!K^I)
        \end{matrix}\right.\\
        &\hspace{8.2em}
        \left.\begin{matrix}
        I \\
        -M^{-1}(K^P\kern-0.4em+\!D^G\kern-0.3em+\!H^{GL}H^{inv}K^{LG})
        \end{matrix}\right]
    \end{aligned}
\end{align}
\begin{equation}
    \label{eq:stateSpaceDLAAFinal}
    \begin{bmatrix}
        \dot{\delta} \\ \dot{\omega}
    \end{bmatrix} = A^*\begin{bmatrix}
        \delta \\ \omega
    \end{bmatrix} +\begin{bmatrix}
        0 \\ M^{-1}H^{GL}H^{inv}
    \end{bmatrix}P^{LS}
\end{equation}
\normalsize

\vspace{-2mm}
\subsection{MDLAA Modeling}
\label{subsec:ModelOfMDLAA}

Based on \cite{MDLAA}, to model the MDLAA, we first need to convert the state-space representation to discrete form, which is shown in \eqref{eq:stateSpaceDiscrete}. We assign $u = p^a \in \mathbb{R}^{|\mathcal{L}|}$, which is the attack vector of the manipulated load. Then, we assign $y = \omega^s \in \mathbb{R}^{|\mathcal{G}|}$ as the vector of frequency measurements at sensor buses. We set the $k^a$ to $1$ at the time step when the attack begins.

\small
\begin{equation}
    \label{eq:stateSpaceDiscrete}
    \begin{aligned}
        x(k^a+1) &= Ax(k^a) + Bu(k^a) \\
        y(k^a) &= Cx(k^a),\quad  k^a = 1,...,k^a_{max}
    \end{aligned}
\end{equation}
\normalsize

Next, the attack is divided into two phases. During the first phase, we collect $T^a$ measurements of frequency and the corresponding attack vectors as vectors $\omega^{sd}_{[1,T^a]}$ and $p^{ad}_{[1,T^a]}$. For an attack to be successful we must ensure that the $p^{ad}$ is persistently excited of order $N^{ap}+T^{ini}+n$ where $T^a \geq (|\mathcal{L}|+1)(T^{ini}+N^{ap}+n)-1$ which allows to create a predictor which based on the past $T^{ini}$ collected samples can estimate the future $N^{ap}$ steps. For the signal to be persistently excited of order of order L, its Hankel matrix must have a full rank. The general form of the Hankel matrix for a signal $x$ of $T^a$ samples is defined in \eqref{eq:hankelMatrix}. The last preparation step is to reshape the collected attack and measurement vectors into appropriate forms. First, we create two Hankel matrices $\mathcal{H}(p^{ad})$ and $\mathcal{H}(\omega^{sd})$ each with the $T^{ini}+N^{ap}$ rows. Then, we separate these matrices into two parts. The first is used to estimate initial conditions, while the second is used to predict future system behavior. The matrices are shown in \eqref{eq:attackVectorMatrix} and \eqref{eq:frequencyMeasurementsMatrix} with subscript $p$ representing the part for initial condition estimation and $f$ the part for system behavior prediction. The $p$ part includes the matrix's first $T^{ini}$ rows, and the $f$ part includes the last $N^{ap}$ rows.

\small
\vspace{-2mm}
\begin{equation}
    \label{eq:hankelMatrix}
    \mathcal{H}_L(x_{[1,T^a]})\! :=\!\! \begin{bmatrix}
        x(1) & \kern-0.6em x(2) & \kern-0.6em... & \kern-0.55em x(T^a\kern-0.45em-\!L\!+\!1) \\
        x(2) & \kern-0.6em x(3) & \kern-0.6em... & \kern-0.55em x(T^a\kern-0.45em-\!L\!+\!2) \\
        \vdots & \kern-0.6em\vdots & \kern-0.6em\ddots & \kern-0.55em\vdots \\
        x(L) & \kern-0.6em x(L+1) & \kern-0.6em... & \kern-0.55em x(T^a)
    \end{bmatrix}
\end{equation}
\vspace{-4mm}
\begin{align}
        \label{eq:attackVectorMatrix}
        \begin{bmatrix}
            P^a_p \\ P^a_f
        \end{bmatrix} &:= \mathcal{H}_{(T^{ini}+N^{ap})}(p^{ad}_{[1,T^a]}) \\
        \label{eq:frequencyMeasurementsMatrix}
        \begin{bmatrix}
            \Omega^s_p \\ \Omega^s_f
        \end{bmatrix} &:= \mathcal{H}_{(T^{ini}+N^{ap})}(\omega^{sd}_{[1,T^a]}) 
\end{align}
\normalsize

As mentioned in \cite{MDLAA}, this method uses the Fundamental Lemma of behavioral system theory, which states that if the $p^a$ is persistently excited of rank $T^{ini}+N^{ap}$, then it is possible to describe all future trajectories of the system can be described by the linear combinations of the Hankel matrices blocks using the predictor vector $g\! \in\! \mathbb{R}^{T^a\!-T^{ini}\!-N^{ap}+1}$ as shown in \eqref{eq:fundamentalLemmaOfBehavioralSystemTheory}. Here, the $p^a_{ini}$ and $\omega^s_{ini}$ are the last $T^{ini}$ collected samples, and $p^a_f$ and $\omega^s_f$ are the predicted attack vectors and frequency values in the next $N^{ap}$ steps. After preparing the collected samples, we can start the online phase of the attack. Here, we are solving the optimization problem shown in \eqref{eq:MDLAAOptimizationProblem} with constraints shown in \eqref{eq:fundamentalLemmaOfBehavioralSystemTheory} to \eqref{eq:MDLAAConstraint3}. We optimize so the predicted frequency $\omega^s_f$ approaches the desired frequency $\omega^r$, which, if reached, indicates the successful attack and $p^a_f$ is as low as possible to limit the resources needed to execute the attack. For compact representation, we define $\|A\|^2_B := A^TBA$. For tuning the attack, we use two weight matrices, $Q$ and $R$, that balance how big priority the optimization assigns to each function component. Finally, using the Algorithm \ref{alg:MDLAAAlgorithm}, we execute the MDLAA.

\small
\begin{equation}
    \label{eq:MDLAAOptimizationProblem}
    \min_{g, p^a_f, \omega^s_f} \sum^{N^{ap}-1}_{t^a=0} \left( \|\omega^s_f(t^a)-\omega^r\|^2_Q + \|p^a_f(t^a)\|^2_R\right)
\end{equation}
\begin{equation}
    \label{eq:fundamentalLemmaOfBehavioralSystemTheory}
    \textrm{s.t.}\quad \begin{bmatrix}
        P^a_p & \kern-0.6em\Omega^s_p & \kern-0.6em P^a_f & \kern-0.6em\Omega^s_f 
    \end{bmatrix}^T g = \begin{bmatrix}
        p^a_{ini} & \kern-0.6em\omega^s_{ini} & \kern-0.6em p^a_f & \kern-0.6em\omega^s_f
    \end{bmatrix}^T
\end{equation}
\begin{equation}
    \label{eq:MDLAAConstraint1}
    p^a_{ini} = [p^a_f(k^a-T^{ini}), ..., p^a_f(k^a-1)]^T
\end{equation}
\begin{equation}
    \label{eq:MDLAAConstraint2}
    \omega^s_{ini} = [\omega^s_f(k^a-T^{ini}), ..., \omega^s_f(k^a-1)]^T
\end{equation}
\begin{equation}
    \label{eq:MDLAAConstraint3}
    |p^a_f(t^a)| \leq |p^{max}(t^a)|, \quad t^a \in \{0,1,...,N^{ap}-1\}
\end{equation}
\vspace{-4mm}
\begin{algorithm}
    \caption{Measurement-based DLAA}
    \label{alg:MDLAAAlgorithm}
    \begin{algorithmic}[1]
        \STATE \textbf{Inputs:} Collected data $[p^{ad}, \omega^{sd}]^\top$
        \STATE \textbf{Output:} Optimal attack vector $p^{a^*}$
        \STATE \textbf{Initialize:} $\omega^r, T^a, k^a_{\text{max}}, p^{\text{max}}, N^{ap}, N^{ac}, R, Q$
        \STATE \textbf{Build:} Hankel matrices $P^a_p, P^a_f, \Omega^s_p, \Omega^s_f$
        \STATE \textbf{Initialize:} input/output data $[p^a_{\text{ini}}, \omega^s_{\text{ini}}]^\top$
        \STATE \textbf{Set:} loop counter $k^a \gets 1$
        \WHILE{$k^a < k^a_{\text{max}}$ \textbf{or} frequency below $\omega^r$}
            \STATE Solve optimization problem \eqref{eq:MDLAAOptimizationProblem}–\eqref{eq:MDLAAConstraint3} for optimal $g^*$
            \STATE Compute optimal attack sequence $p^{a^*} = P^a_f g^*$
            \STATE Apply attack inputs $(p^a_f(k^a), \ldots, p^a_f(k^a+N^{ac})) = (p^{a^*}(0), \ldots, p^{a^*}(N^{ac}))$
            \FOR{$N^{ac} \leq N^{ap} - 1$}
                \STATE Update loop counter: $k^a \gets k^a + N^{ac}$
                \STATE Update $p^a_{\text{ini}}$ and $\omega^s_{\text{ini}}$ with latest attack vectors and measurements
            \ENDFOR
        \ENDWHILE
    \end{algorithmic}
\end{algorithm}
\vspace{-2mm}
\normalsize

\vspace{-2mm}
\subsection{LAA effect on LFC}
\label{subsec:ConncetionOfModelsOfLFCAndLAA}
\vspace{-1mm}

For LFC, we consider LAA from a power flow perspective \cite{LAALFCTolerantFramework}, as shown in \eqref{eq:generalLAA}. The $\mathcal{N}_e$ is the set of vulnerable load buses. The $P$ is the original power, $U$ represents the bus voltage magnitude, and $\theta_{ij}$ is the phase angle difference between two buses. The $G_{ij}$ and $B_{ij}$ are the real and imaginary parts of the admittance between two buses, and $d$ is the load altered by LAA. Buses directly connected to bus $i$ are indexed by $j$.

\vspace{-4mm}
\small
\begin{equation}
    \label{eq:generalLAA}
    P_{is} + d = U_i \sum_{j\in\mathcal{N}_i} U_j (G_{ij}cos(\theta_{ij}) + B_{ij}sin(\theta_{ij}), \forall i \in \mathcal{N}_e
\end{equation}
\normalsize
\vspace{-3mm}

The general state-space representation can be expressed by \eqref{eq:generalLAALFC}. However, the power flow equations must keep the form of \eqref{eq:generalLAA}. The $x$ is the system state, $y$ is the system output, and $u$ is the system input. The $f$ and $g$ are algebraic functions, while $d$ represents the LAA alterations. For SLAA, $d$ affects the $u$, while for DLAA, $d$ influences the system matrix within $f$.

\vspace{-2mm}
\small
\begin{equation}
    \label{eq:generalLAALFC}
    \dot{x} = f(x, u, d);\qquad
    y = g(x)
\end{equation}
\normalsize
\vspace{-3mm}

%% file: 3_implementation.tex
\vspace{-4mm}
\section{Implementation}
\label{sec:Implementation}
\vspace{-1mm}

\begin{figure}[t]
    \centering
    \vspace{-2mm}
    \includegraphics[width=0.9\linewidth]{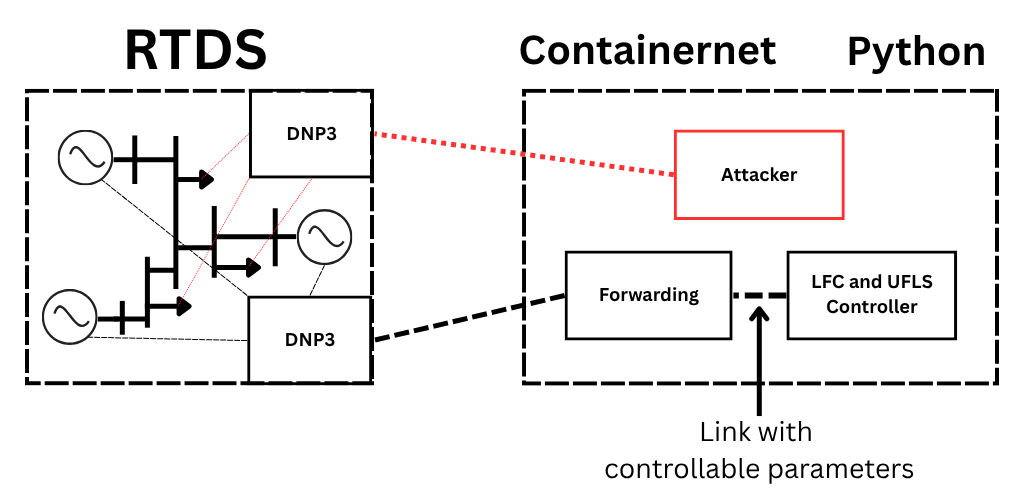}
    \vspace{-3mm}
    \caption{Co-simulation architecture overview.}
    \vspace{-7mm}
    \label{fig:architecture}
\end{figure}

The co-simulation environment developed in this work integrates RTDS, a real-time digital simulator, with Containernet, a communication network emulator. The architecture, shown in Fig. \ref{fig:architecture}, consists of two main components: a real-time power system simulation, including generation, load, and DNP3 outstations, and an emulated communication network responsible for calculating and coordinating control actions and adversarial behavior. This setup allows tight coupling of physical system states and cyber-layer traffic, enabling attack scenarios that depend on real-time feedback from the grid, such as DLAAs.

The power system model is implemented on the RTDS platform using RSCAD. It includes an implementation of the IEEE 39-bus system with three LFC areas, generator and load models, and an UFLS mechanism. Each control area consists of multiple generators with turbine and governor dynamics. Frequency deviations are sent to and processed by the LFC and UFLS Controller in the network part. The Area Control Error (ACE) is computed based on local frequency and tie-line flows and is used to drive LFC logic. The model also incorporates a configurable attack injection interface, allowing real-time load changes to be triggered from the cyber domain. DLAA logic is implemented using negative feedback on measured frequency, while MDLAA uses externally supplied attack vectors computed based on the constructed predictor.

The communication network is emulated using Containernet, a fork of Mininet that natively supports Docker containers as network hosts. This framework enables realistic network conditions, control scenarios, and the execution of attack coordination logic. Networked hosts include implementations of attack control logic and controllers for grid control and protection mechanisms. LAAs are launched from the attacker container based on received frequency data. The internal link between Forwarding and LFC and UFLS Controller can alter the latency and packet loss to replicate adverse communication conditions. All cyber operations, including attacks, operate in real-time and interact with the physical layer via standardized protocol DNP3, ensuring synchronization with the RTDS simulator.

A key feature of the environment is its flexibility and modularity. Each simulation component, like generators, controllers, attack agents, and communication nodes, can be independently modified or extended. The co-simulation setup supports adding new power system models, including different topologies or control schemes and new classes of attacks. Researchers can evaluate the effects of various network degradations or test the defensive mechanisms. The system’s modular architecture enables iterative development and scalable experimenting, making it a practical tool for cybersecurity research in power grids.

The entire co-simulation platform is open-source and designed for reproducibility. RSCAD power system models and emulation code are provided for deploying and coupling the emulated network with RTDS. This allows other researchers to replicate the presented scenarios or develop new test cases, supporting future cyber-physical power system security work \cite{RTDSModelsGithub, CoSimulationGithub}.

%% file: 4_results.tex
\vspace{-2mm}
\section{Co-Simulation results}
\label{sec:Co-SimulationResults}
\vspace{-1.5mm}

After model definitions and implementation details, we examine the impact of LAAs and network conditions on LFC and system stability through co-simulation. We analyze several attack scenarios, observing how LAAs influence frequency response and stability. Then, we observe how the control mechanisms such as LFC and UFLS help to mitigate the effects of attacks. Finally, we examine how the altered network conditions influence the effectiveness of control mechanisms.

To evaluate the impact of LAAs on the system, we simulated six attack scenarios. The attack always begins $30$ seconds after the simulation starts to exclude any initial setup disturbances. Each scenario includes variations involving activation of LFC and UFLS and, when relevant, manipulation of network conditions such as delay or packet loss. For the UFLS, the activation thresholds adhere to ones defined in \cite{UFLSRequirements}. Table \ref{tab:UFLSThresholds} shows implemented UFLS thresholds and percent of shedded load. By default, each generator implements the primary control.

The frequency plots, include three pairs of lines indicating the thresholds for plant and apparatus operation requirements, as in the Saudi Arabia Grid Code \cite{saudiGrid}. The generator plants and apparatus are designed to operate within a frequency range of $57.0$ Hz to $62.5$ Hz. We consider an attack successful if the frequency exceeds this range or maintains it within defined thresholds for longer than the specified operational limits. All frequency thresholds are presented in Table \ref{tab:SaudiThresholds}.

{
\setlength{\tabcolsep}{4pt}
\begin{table}[t]
    \centering
    \caption{UFLS thresholds and percent of shedded load}
    \vspace{-2mm}
    \begin{tabular}{||c|c|c|c||}
        \hline\hline
        \begin{tabular}{c} \textbf{UFLS} \\ \textbf{stage} \end{tabular} & \begin{tabular}{c} \textbf{Frequency} \\ \textbf{threshold (Hz)} \end{tabular} & \begin{tabular}{c} \textbf{Shedded} \\ \textbf{load (\%)}\end{tabular} & \begin{tabular}{c} \textbf{Cumulative} \\ \textbf{shedded load (\%)} \end{tabular} \\ \hline
        Stage 1 & 59.5 & 7.0 & 7.0 \\ \hline
        Stage 2 & 59.3 & 7.0 & 14.0 \\ \hline
        Stage 3 & 59.1 & 7.0 & 21.0 \\ \hline
        Stage 4 & 58.9 & 7.0 & 28.0 \\ \hline\hline
    \end{tabular}
    \vspace{-3mm}
    \label{tab:UFLSThresholds}
\end{table}
}

\begin{table}[t]
    \centering
        \caption{Saudi Arabia grid code frequency thresholds \cite{saudiGrid}.}
        \vspace{-2mm}
    \begin{tabular}{||c|c|c||}
        \hline\hline
        \textbf{\begin{tabular}{c} Below Nominal \\ Frequency [Hz]\end{tabular}} & \textbf{\begin{tabular}{c} Above Nominal \\ Frequency [Hz]\end{tabular}} & \textbf{\begin{tabular}{c} Operation \\ Requirement\end{tabular}} \\
        \hline
        58.8 - 60.0 & 60.0 - 60.5 & Continuous \\
        \hline
        57.5 - 58.7 & 60.6 - 61.5 & For 30 minutes \\
        \hline
        57.0 - 57.4 & 61.6 - 62.5 & For 30 seconds \\
        \hline\hline
    \end{tabular}
    \vspace{-6mm}
    \label{tab:SaudiThresholds}
\end{table}

\begin{figure*}
    \begin{subfigure}[t]{0.32\textwidth}
        \centering
        \includegraphics[width=0.65\linewidth]{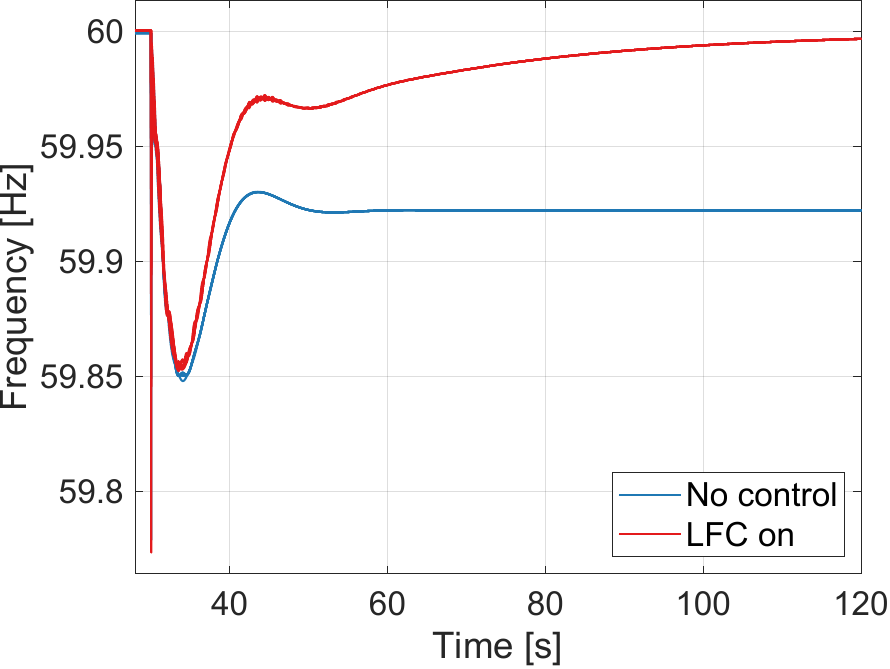}
        \vspace{-2mm}
        \caption{Scenario I - SLAA, variants 1 and 2.}
        \label{fig:SLAALFC}
    \end{subfigure}
    \begin{subfigure}[t]{0.32\textwidth}
        \centering
        \includegraphics[width=0.65\linewidth]{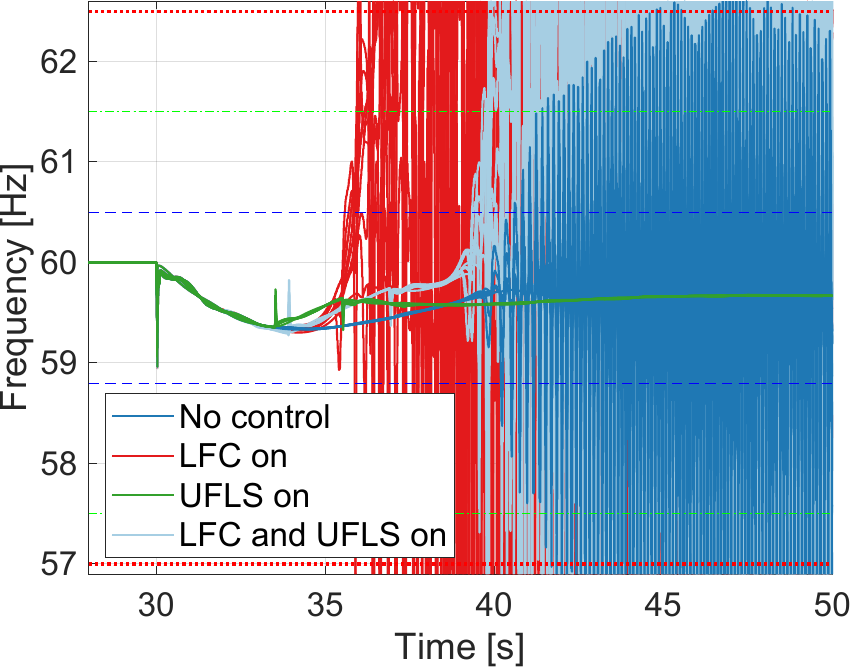}
        \vspace{-2mm}
        \caption{Scenario II - SLAA, variants 1 to 4.}        
        \label{fig:SLAABigLFCUFLS}
    \end{subfigure}
    \begin{subfigure}[t]{0.32\textwidth}
        \centering
        \includegraphics[width=0.65\linewidth]{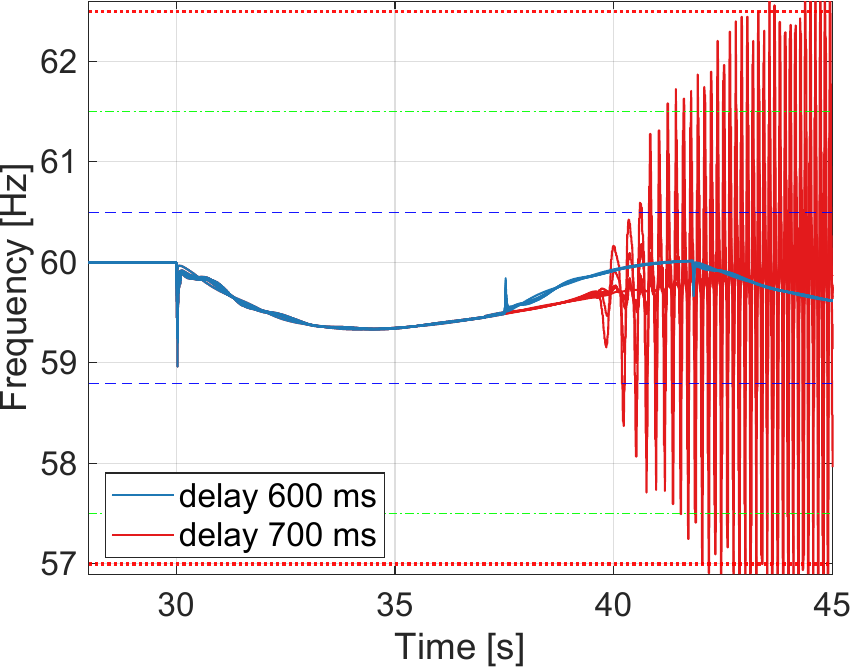}
        \vspace{-2mm}
        \caption{Scenario II - SLAA, variants 5 and 6.}
        \label{fig:SLAABigUFLSDelays}
    \end{subfigure}
    \vspace{-2mm}
\caption{Frequency responses for scenarios I and II - SLAA.}
\vspace{-4mm}
\label{fig:SLAAResults}
\end{figure*}

\begin{figure*}
    \begin{subfigure}[t]{0.24\textwidth}
        \centering
        \includegraphics[width=0.85\linewidth]{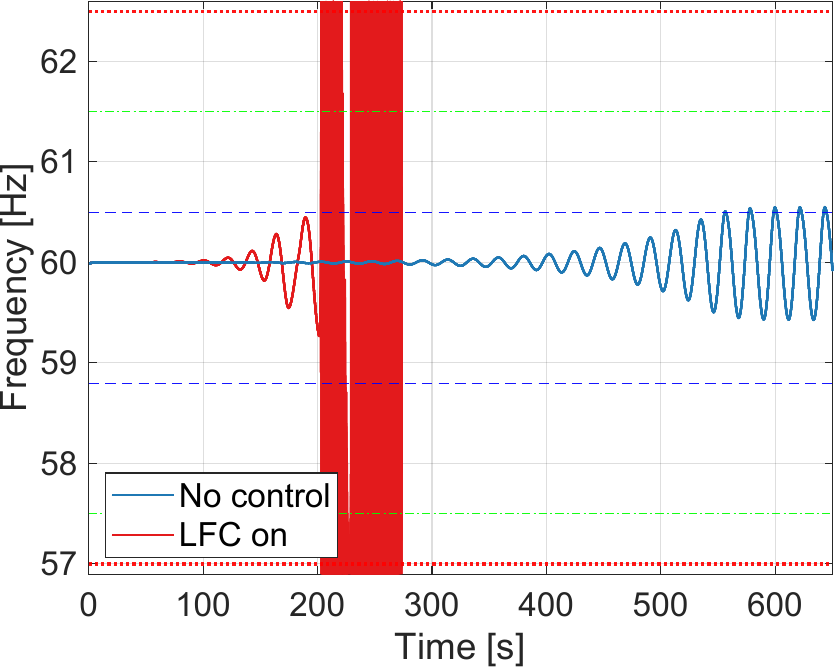}
        \vspace{-2mm}
        \caption{Scenario III - DLAA, variants 1 and 2.}        
        \label{fig:DLAALFC}
    \end{subfigure}
    \begin{subfigure}[t]{0.24\textwidth}
        \centering
        \includegraphics[width=0.85\linewidth]{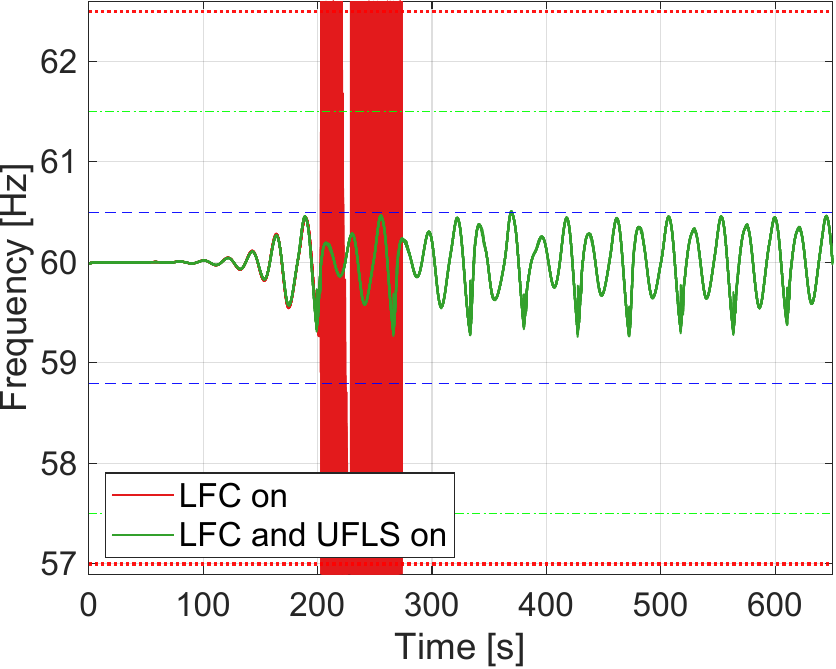}
        \vspace{-2mm}
        \caption{Scenario III - DLAA, variants 2 and 3.}        
        \label{fig:DLAALFCUFLS}
    \end{subfigure}
    \begin{subfigure}[t]{0.24\textwidth}
        \centering
        \includegraphics[width=0.85\linewidth]{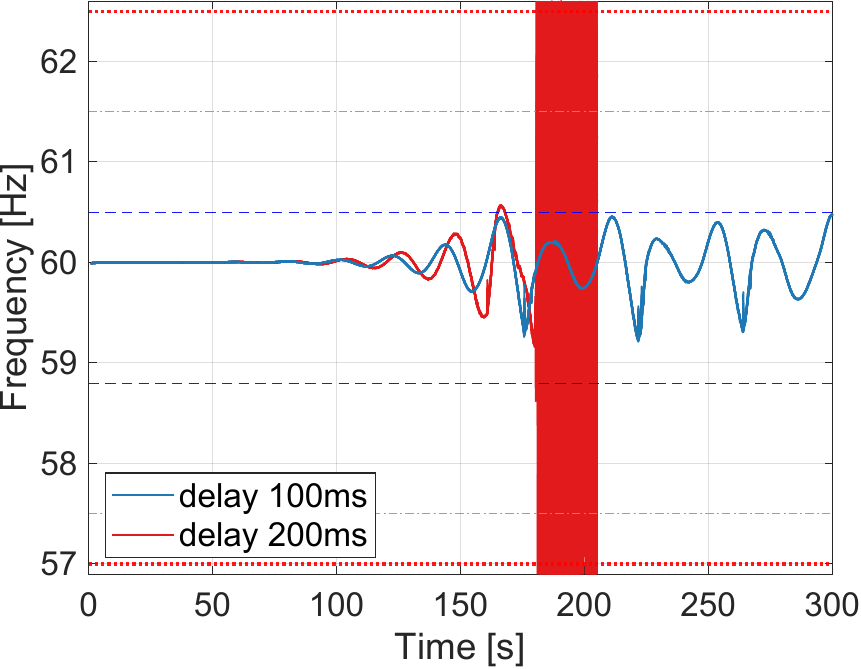}
        \vspace{-2mm}
        \caption{Scenario III - DLAA, variants 4 and 5.}
        \label{fig:DLAALFCUFLSDelays}
    \end{subfigure}
    \begin{subfigure}[t]{0.24\textwidth}
        \centering
        \includegraphics[width=0.85\linewidth]{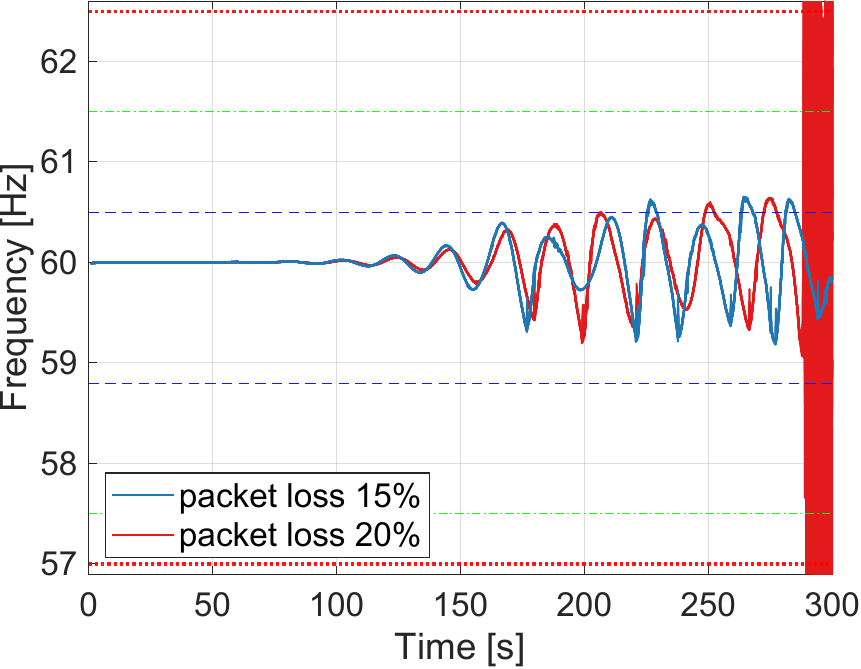}
        \vspace{-2mm}
        \caption{Scenario III - DLAA, variants 6 and 7.}
        \label{fig:DLAALFCUFLSLosses}
    \end{subfigure}

    \hfill
    
    \begin{subfigure}[t]{0.24\textwidth}
        \centering
        \includegraphics[width=0.85\linewidth]{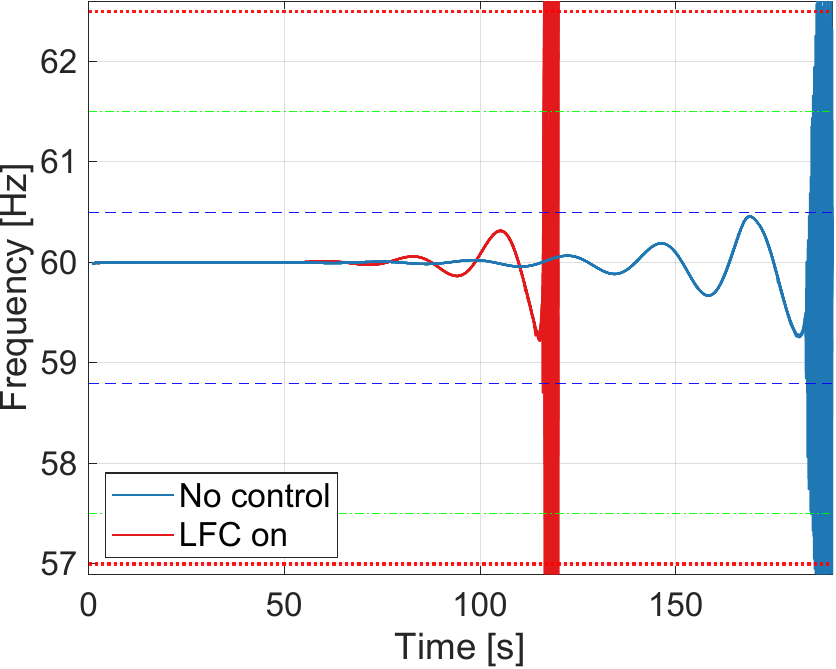}
        \vspace{-2mm}
        \caption{Scenario IV - DLAA, variants 1 and 2.}
        \label{fig:DLAABigLFC}
    \end{subfigure}
    \begin{subfigure}[t]{0.24\textwidth}
        \centering
        \includegraphics[width=0.85\linewidth]{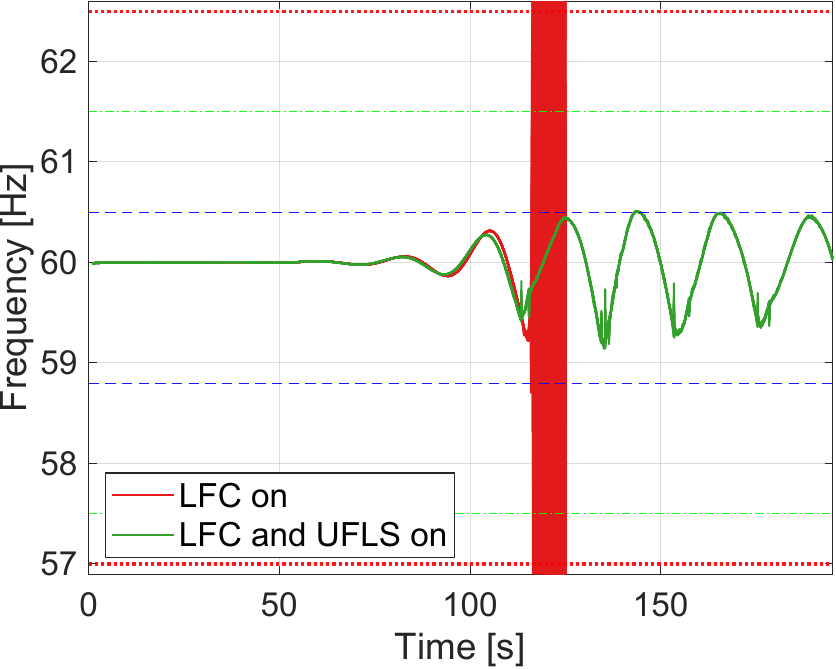}
        \vspace{-2mm}
        \caption{Scenario IV - DLAA, variants 2 and 3.}
        \label{fig:DLAABigLFCUFLS}
    \end{subfigure}
    \begin{subfigure}[t]{0.24\textwidth}
        \centering
        \includegraphics[width=0.85\linewidth]{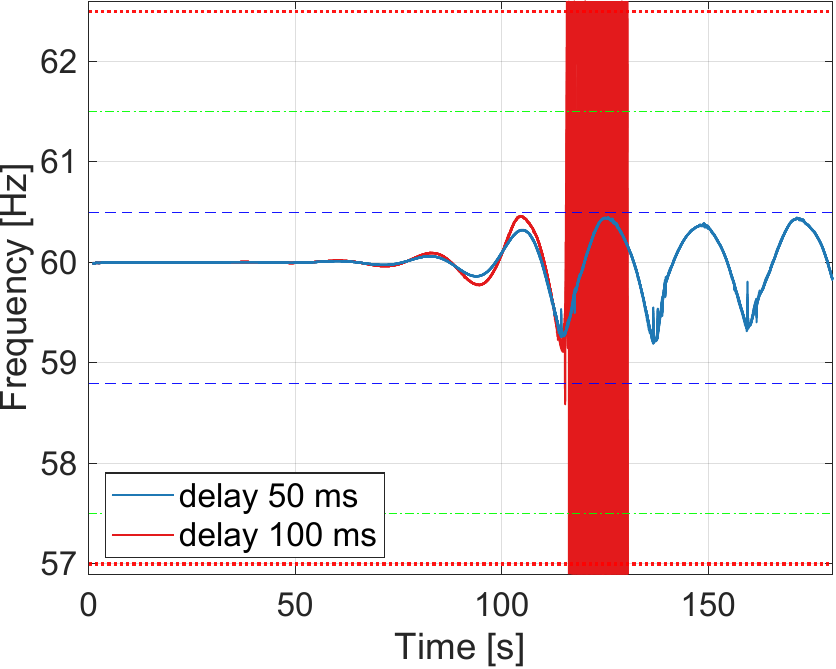}
        \vspace{-2mm}
        \caption{Scenario IV - DLAA, variants 4 and 5.}
        \label{fig:DLAABigLFCUFLSDelays}
    \end{subfigure}
    \begin{subfigure}[t]{0.24\textwidth}
        \centering
        \includegraphics[width=0.85\linewidth]{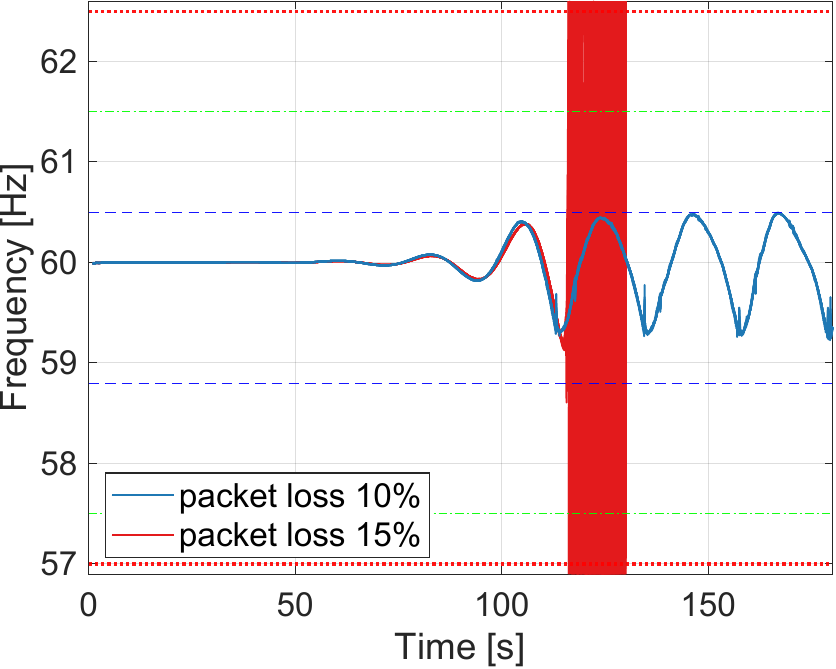}
        \vspace{-2mm}
        \caption{Scenario IV - DLAA, variants 6 and 7.}
        \label{fig:DLAABigLFCUFLSLosses}
    \end{subfigure}
    \vspace{-2mm}
\caption{Frequency responses for scenarios III and IV - DLAA.}
\vspace{-4mm}
\label{fig:ScenarioIResults}
\end{figure*}
    
\begin{figure*}
    \begin{subfigure}[t]{0.32\textwidth}
        \centering
        \includegraphics[width=0.65\linewidth]{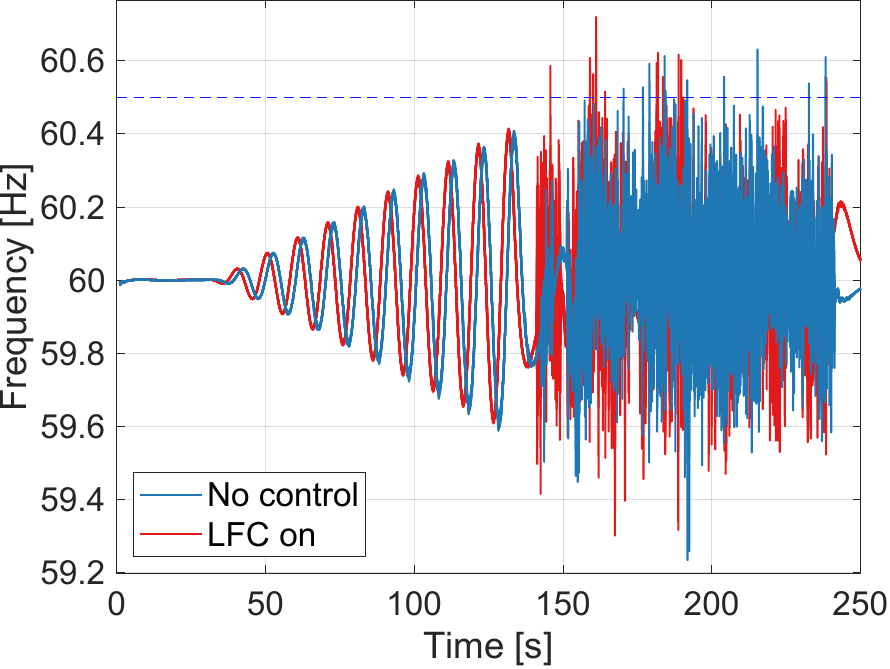}
        \vspace{-2mm}
        \caption{Scenario V - MDLAA, variants 1 and 2.}
        \label{fig:MDLAALFC}
    \end{subfigure}
    \begin{subfigure}[t]{0.32\textwidth}
        \centering
        \includegraphics[width=0.65\linewidth]{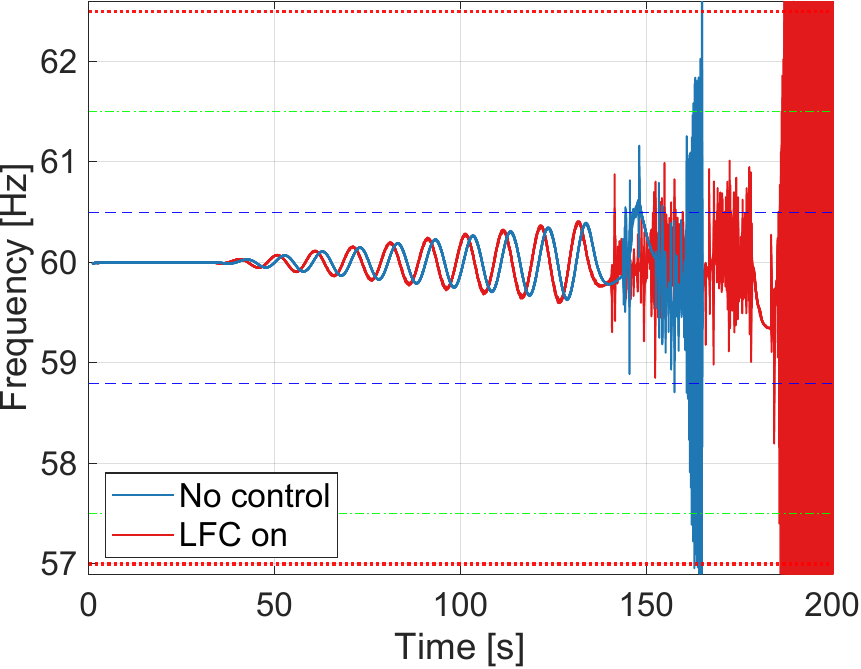}
        \vspace{-2mm}
        \caption{Scenario VI - MDLAA, variants 1 and 2.}
        \label{fig:MDLAABigLFC}
    \end{subfigure}
    \begin{subfigure}[t]{0.32\textwidth}
        \centering
        \includegraphics[width=0.65\linewidth]{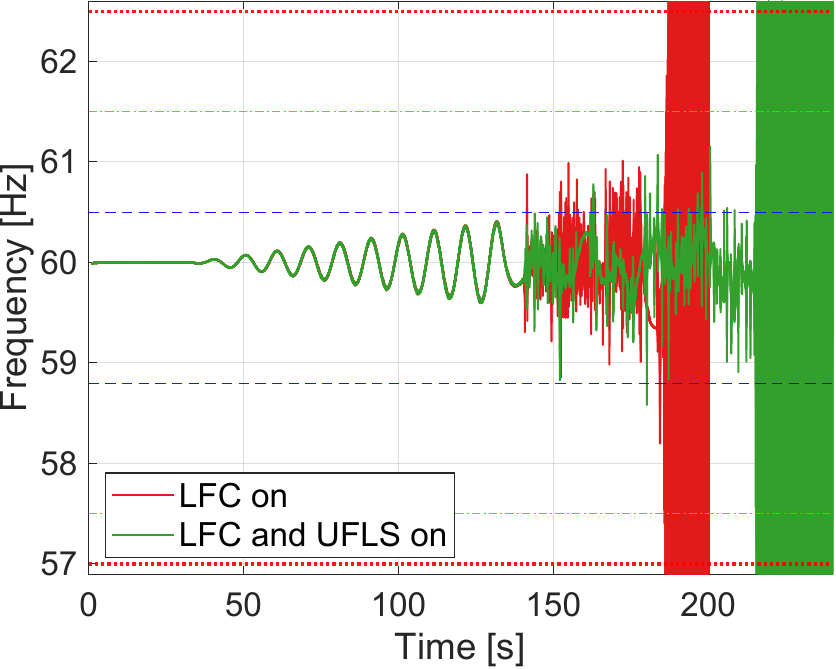}
        \vspace{-2mm}
        \caption{Scenario VI - MDLAA, variants 2 and 3.}
        \label{fig:MDLAABigLFCUFLS}
    \end{subfigure}
    \vspace{-2mm}
\caption{Frequency responses for scenarios V and VI - MDLAA.}
\vspace{-6mm}
\label{fig:ScenarioIIResults}
\end{figure*}

The attack is launched in the first four scenarios at load buses 4 and 20, because a single point attack was unsuccessful in all scenarios. Also, they are among the buses with the largest load in the system, making them sufficient to destabilize the frequency and limit the number of attack points. Finally, they are located in two different LFC areas, allowing for a better observation of the LFC dynamics than in a single-area attack. For two last scenarios MDLAA is launched on all load buses.

In \textbf{Scenario I} (Fig. \ref{fig:SLAALFC}), we simulate an SLAA that increases the load on two buses by 20\%. However, it is too weak to destabilize the system. In \textbf{I.1}, we can see that even though the system remains stable, there is a steady state frequency error, which the primary control could not remove independently. Next, in \textbf{I.2} we activate LFC, and as expected, its presence restores the nominal frequency over time. 

In \textbf{Scenario II} (Figs. \ref{fig:SLAABigLFCUFLS} and \ref{fig:SLAABigUFLSDelays}), the SLAA increases the load on buses 4 and 20 by 100\% and 76\%. It is strong enough to disrupt the system when LFC and UFLS are inactive (\textbf{II.1}). In \textbf{II.2}, the LFC is activated. However, in the presence of strong SLAA, it cannot restore the frequency. Instead, the system becomes unstable even faster. In \textbf{II.3}, we activate UFLS instead. After applying the load shedding, the primary control stabilized the system, although keeping an off-nominal frequency level due to the lack of LFC. Variation \textbf{II.4} combines LFC and UFLS, but the system operation is disrupted again. Finally, variations \textbf{II.5} and \textbf{II.6} applies delays on UFLS. When the delay reaches a certain level, UFLS cannot issue the shedding command before the system becomes unstable.

\textbf{Scenario III} (Figs. \ref{fig:DLAALFC}--\ref{fig:DLAALFCUFLSLosses}) presents DLAA with $K^{\!LG}\!\!=\!70$ on both buses, which was too weak to destabilize the system. Initially, in \textbf{III.1}, the attack slowly increases the frequency oscillations. However, the alterations were too slow, and the system adjusted fast enough to remain in a stable oscillating pattern. In variation \textbf{III.2}, after introducing LFC, the oscillations gain progressed significantly faster than in \textbf{III.1}. Eventually, it brought the system to instability around 170 seconds after the launch of the attack. Variation \textbf{III.3} combines LFC and UFLS, which were able to stabilize the system. However, the oscillations were not dampened, but instead, similarly to variation \textbf{III.1}, the system kept oscillating, remaining in the allowable range. When network delay or packet loss were added to LFC and UFLS in variations \textbf{III.4}-\textbf{III.7}, the UFLS could not react in time to prevent the instability.

In \textbf{Scenario IV} (Figs. \ref{fig:DLAABigLFC}--\ref{fig:DLAABigLFCUFLSLosses}), DLAA with $K^{\!LG}\!\!=\!80$ on both buses destabilized the system. Variation \textbf{IV.1} shows that the system with only primary control could not adjust to the attack, leading to instability. Variation \textbf{IV.2} shows that LFC increased the attack efficiency, making the attack successful about 60 seconds earlier. Similarly to \textbf{III.3}, the variation \textbf{IV.3} shows that UFLS can stabilize the system. However, with the constant frequency oscillations. The variations \textbf{IV.4}-\textbf{IV.7} disrupt the UFLS operation, but compared to \textbf{scenario III}, the network conditions requirements are slightly higher. With the increasing attack strength, the margin for error in UFLS shrinks, imposing higher quality requirements on the communication links.

For the \textbf{Scenario V} (Fig. \ref{fig:MDLAALFC}), we see the effect of MDLAA with 30\% of system load, insufficient to destabilize the system. In Fig. \ref{fig:MDLAALFC}, we can see two attack phases. First, the data collection phase with a predefined trajectory, and then the online phase based on predictions. The system remains stable in both variations \textbf{V.1} and \textbf{V.2}. The presence of LFC does not seem to impact the attack result significantly. However, we can notice that LFC's presence slightly slowed the oscillation growth during the offline phase. It contrasts what we observed for \textbf{scenario III} and \textbf{scenario IV}, where the LFC accelerated the oscillation growth. However, in the MDLAA offline phase, the load follows a predefined sinusoidal pattern, while in DLAA, the load reacts to frequency changes. It makes the DLAA exploit the LFC response to accelerate the attack. The lack of instability in this scenario makes the comparison of default variation and activated LFC sufficient.

In \textbf{Scenario VI} (Figs. \ref{fig:MDLAABigLFC} and \ref{fig:MDLAABigLFCUFLS}), the MDLAA is assigned 60\% of system load and can cause system disruption. In variation \textbf{VI.1} we see the system becoming unstable just after the start of the online phase. With LFC active in variation \textbf{VI.2} the attack is again successful, although it became slightly delayed due to the LFC presence. In variation \textbf{VI.3}, both LFC and UFLS are active, which results in an even greater delay before the system disruption. It shows that in this scenario, each added control and security layer improves system robustness as expected. These results suggest that MDLAA, contrary to the DLAA, does not benefit from the presence of LFC. However, it can destabilize the system even with both protections activated, making the analysis of altered network conditions uninsightful.

\vspace{-1mm}

%% file: 5_conclusion.tex
\vspace{-3mm}
\section{Conclusions}
\label{sec:Conclusions}
\vspace{-2mm}

This work presents a modular, open-source co-simulation framework integrating RTDS (power system simulation) with Containernet (network emulation), enabling realistic cyber-physical studies of LAAs on LFC. The study reveals nuanced system responses and adversary interactions by implementing and analyzing SLAA, DLAA, and MDLAA  scenarios under varying network conditions and LFC and UFLS protection mechanisms. Notably, DLAA exploits LFC feedback to accelerate instability, while MDLAA maintains robustness against LFC and can still disrupt operation despite protective measures. The results highlight the critical role of communication reliability and coordination between control layers in enhancing grid resilience against evolving cyber-physical threats.

%% file: acknowledgments.tex
\section*{Acknowledgments}
\vspace{-2mm}
This publication is based upon
work supported by King Abdullah University of Science and Technology under Award No. ORFS-2022-CRG11-5021.